\documentclass[twocolumn,showpacs,preprintnumbers,amsmath,amssymb]{revtex4-1}

\usepackage{graphicx}
\usepackage{dcolumn}
\usepackage{bm}
\usepackage{amsmath}
\usepackage{color}
\usepackage{soul}

\begin{document}

\title{Tamm plasmon photonic crystals : from bandgap engineering to defect cavity}

\author{L. Ferrier$^1$}
\email{lydie.ferrier@insa-lyon.fr}
\author{H.S. Nguyen$^2$}
\author{C. Jamois$^1$}
\author{L. Berguiga$^1$}
\author{C. Symonds$^3$}
\author{J. Bellessa$^3$}
\author{T. Benyattou$^1$}

\affiliation{$^1$Universit\'e de Lyon, Institut des Nanotechnologies de Lyon, INL/CNRS, INSA-Lyon,  7 avenue Jean Capelle, 69621 Villeurbanne, France}
\affiliation{$^2$Universit\'e de Lyon, Institut des Nanotechnologies de Lyon, INL/CNRS, Ecole Centrale de Lyon,  36 avenue Guy de Collongue, 69130 Ecully, France}
\affiliation{$^3$Universit\'e de Lyon, Universit\'e Claude Bernard Lyon 1, CNRS, Institut Lum\`ere Mati\`ere, F-69622, Lyon, France}

\date{\today}
\pacs{}

\begin{abstract}
We report for the first time the bandgap engineering of Tamm plasmon photonic crystals - Tamm plasmon structures of which the metallic layer is periodically patterned into lattice of subwavelength period.  By adopting a double period design, we evidenced experimentally a complete photonic bandgap up to $150\,nm$ in the telecom wavelength range. Moreover, such design offers a great flexibility to tailor on-demand, and independently, the band-gap size from $30\,nm$ to $150\,nm$ and its spectral position within $50\,nm$. Finally, by implementing a defect cavity within the Tamm plasmon photonic crystal, an ultimate cavity of $1.6\mu m$ supporting a single highly confined Tamm mode is experimentally demonstrated. All experimental results are in perfect agreement with numerical calculations. Our results suggests the possibility to engineer novel band dispersion with surface modes of hybrid metallic/dielectric structures, thus open  the way to Tamm plasmon towards applications in topological photonics, metamaterials and parity symmetry physics.

\end{abstract}

\maketitle

Photonic bandgap engineering plays a key role in modern photonics since it allows an ultimate control of photon propagation in periodic dielectric or metallic media~\cite{Joannopoulos1997}. Similarly to the forbidden gaps of electrons in crystals, photons cannot propagate in the direction of periodicity for a given range of energies. The simplest example of the apparition of a photonic bandgap in a periodic layered media is the well know Bragg mirror~\cite{PochiYeh2005}. At the edges of the forbidden gap, as the group velocity is close to zero, the high density of states allows to strongly enhance the light matter interaction. This unique property has been widely used for the realization of various nanophotonic devices such as integrated and mirror free microlasers~\cite{Ryu2002} or enhancing non linear effects~\cite{Monat2009}. Modifying locally the periodic pattern of a photonic crystal leads to localized states within the photonic bandgap. This modification can take the form of a line~\cite{Foresi1997,Kotlyar2004} or a point defect~\cite{Painter1999,Qi2004,Akahane2003} within the periodic modulation, or of a photonic heterostructure~\cite{Song2005,Istrate2006,Ferrier2008}. The concept of photonic nanocavity has been widely used to demonstrate lasing in microcavities~\cite{O.Painter1999,Altug2006}, enhancement of spontaneous emission~\cite{Boroditsky1999,Englund2005}, strong coupling regime between quantum emitters and photonic defect cavity~\cite{Christ2003,Yoshle2004}, and exciton polariton lasing in photonic crystal cavities~\cite{Azzini2011}.

Apart from all dielectric or all metallic structures, Tamm plasmon modes are optical states localized at the interface of a dielectric Bragg mirror and a thin metal layer~\cite{Kaliteevski2007}, thus inheriting properties of both plasmons and cavity modes. Unlike conventional plasmons, they present much less losses and their parabolic dispersion enable a coupling to free space modes~\cite{Sasin2008}. One of the main advantage of Tamm plasmons is the tailoring of the field confinement only by patterning the thin metal layer, avoiding deep etching of nanostructures. Up to now, lateral mode confinement has been achieved by defining micrometric metallic structures such as disks~\cite{Gazzano2011,Symonds2013,Braun2015} and rectangles~\cite{Lheureux2015} leading to three dimensional confinement of light experimentally evidenced by a set of discrete Tamm modes in energy. Br\"{u}ckner et al.~\cite{Bruckner2012} have shown recently that thin metallic periodic arrays with period of a few microns replicate the parabolic dispersion of Tamm modes in k-space. At the edges of the Brillouin zone, where $k=\pm m\pi/a$ ($m$ being an integer) the local density of state is greatly enhanced due to the crossing of the replicated dispersions. However, as the grating period was ten times $a/\lambda$, this device works in the Bragg scattering regime. In this regime, the bandgap if existing is much smaller than the spectral width of the modes, making it impossible to realize photonic bandgap and microcavity Tamm structures. As consequence, new metallic nanostructures need to be designed To gain more versatility in the dispersion tailoring of optical Tamm modes and ultimate light confinement within photonic Tamm cavities.
\begin{figure*}[htb]
\begin{center}
\includegraphics[width=17cm]{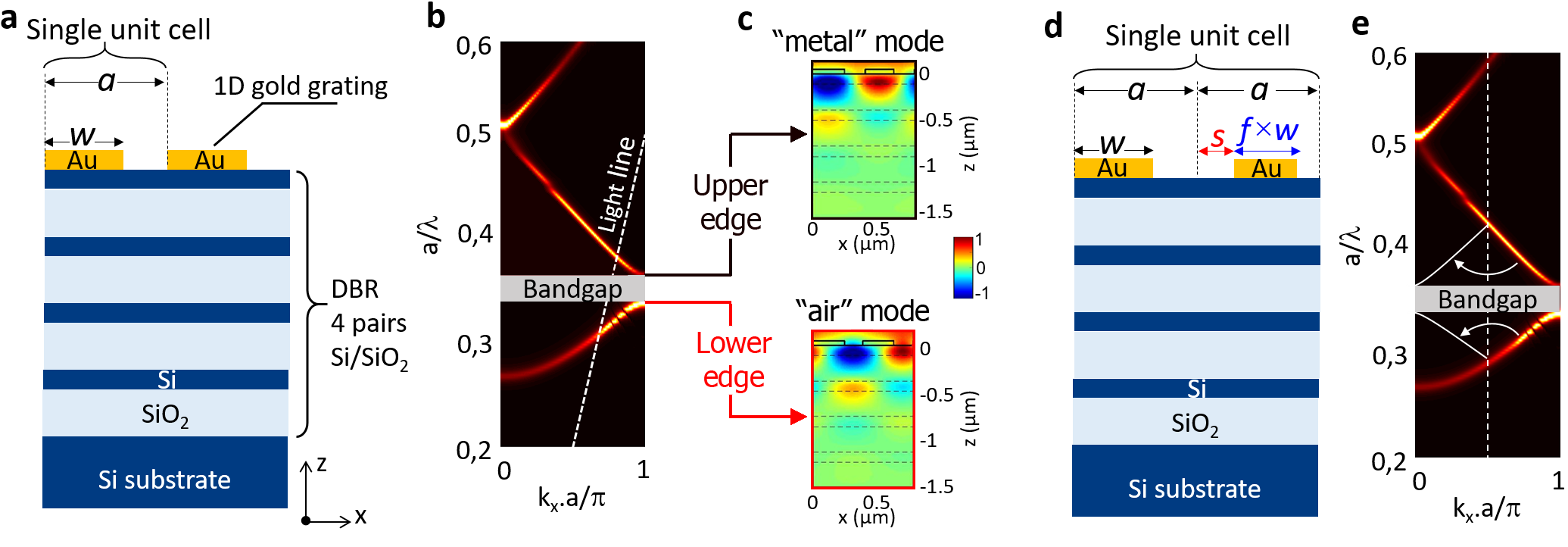}
\end{center}
\caption{\label{fig1}{(a) Sketch of the photonic crystal Tamm plasmon structure. (b) Numerical simulation of the energy-momentum dispersion diagram of a photonic crystal Tamm plasmon with 50nm layer of gold, the period of the pattern is $a=400\,nm$ and the width of the metal stripes is $w=250\,nm$. (c) Calculated electric field distribution of two band-edge modes. (d) Sketch of the double period design. (e) Band-folding in double period design: band-edge modes at X point ($k_x=\pi/a$) are brought to the $\Gamma$ point ($k_x=0$).}}
\end{figure*}

In this Letter, we show that a complete photonic bandgap can be opened by patterning the metallic layer of Tamm plasmon structures into periodical lattice in the photonic crystal regime (i.e. $a/\lambda\ll1$). Moreover, an original design, named ``double periodic'', offers an ``on-demand'' control of the bandgap size and bandgap spectral position. Angle-resolved measurements reveal the opening of a photonic bandgap as large as $150~nm$ around $1.5\mu m$. The bandgap size and the bandgap position can be freely and independently tuned within the telecom range. Experimental Tamm dispersions show perfect agreement with numerical calculations. To highlight the potential of bandgap engineering, we  confine Tamm modes laterally by introducing a defect cavity of only $1.6\mu m$ within the bandgap.  These experimental results, where the light confinement completely rely on the metal patterning, would open the way to all photonic bandgap applications on a hybrid metal/dielectric platform.

We first highlight the mechanism of bandgap opening of Tamm plasmon photonic crystals (i.e. Tamm structures with periodic metallic layer in the regime $a/\lambda\ll1$). Figure 1(a) shows the schematic of the photonic Tamm structure consisting of a 1D periodic gold grating deposited onto a silicon-based Distributed Bragg Reflector (DBR) (consisted of 4 pairs $\lambda/4$ of Si/SiO$_2$ with centre wavelength $\lambda_{Bragg}=1.5\,\mu m$, leading to a stopband from $1.2\mu m$ to $2\mu m$). The energy-momentum dispersion of TE-polarized Tamm modes, numerically simulated by  Finite-Difference Time-Domain method (FDTD) method, is presented in Fig.~1(b). It shows clearly a complete photonic bandgap at the edge of the first Brillouin zone (i.e. X point with $k_x=\pi/a$). This bandgap openning is resulted from the diffractive coupling between the forward and backward Tamm modes propagating beneath the metallic layer. Impressively, the bandgap-size of $120\,nm$, corresponding to $15\%$ of the stopband of the DBR, is multiple times larger than the linewidth of Tamm modes ($\sim 10\,nm$). Figure 1(c) depicts the electric field distribution at the two band edges. We note that: i)The antinodes of the upper-edge mode are located beneath  the metal stripes, and we will refer to such localization as ``metal'' mode. ii) The antinodes of the lower-edge mode are located beneath  the air groove. We will refer to such localization as ``air'' mode. The fact that the ``metal'' mode is at higher energy than the ``air'' mode can be explained by its tighter confinement. Indeed, as shown in Fig.~1(b), contrary to the ``metal'' mode which is totally localized in the Bragg, the ``air'' mode penetrates slightly out of the Bragg.

As discussed previously, an important bandgap will be opened when the metal layer of a Tamm structure is periodically patterned with a subwavelength period. However, this design is not flexible for bandgap engineering. Moreover,  the band-edges are located below the light line, thus they are not experimentally accessible from farfield measurements such as reflectivity, transmission and photoluminescence experiments. To overcome these drawbacks, we adopt in this work a ``double period'' design with two gold stripes per unit cell [see Fig.~1(d)] instead of one as in the initial ``single period'' [see Fig.~1(a)]. This ``double period'' design is obtained by shifting and shrinking the second gold stripe with respect to its initial position and size. Quantitatively, the period of single unit cell is doubled to $2a$ with the second gold stripe shifted by a distance $s$ ($0<s/a<1$) and shrunk by a factor $f$ ($0<f<1$). The ``double period'' offer a twofold advantage: \textbf{i) Tamm modes at the band-edges are now directly accessible from free space and can be probed via farfield experiments}. Indeed, the Brillouin zone gets twice smaller, thus the band-edges initially located below the light line at $X$-point ($k_x=\pi/a$) will be folded to $\Gamma$-point ($k_x=0$) [see figure 1 (e)]~\cite{Nguyen2018}.  \textbf{ii) The shift parameter $s$ and the shrinking parameter $f$ provide efficient degrees of freedom for bandgap engineering.} Indeed, by implementing a shift or a shrinking of the gold stripes, the confinement of the band-edge modes would be greatly modified, leading to a change of the bandgap size or/and the bandgap position.

Our sample is fabricated on a silicon platform to take advantage of the high index contrast between silica and silicon for the realization of the DBR. As mentioned previously, the DBR is constituted by 4 pairs of Si/SiO$_2$ on silicon substrate, centred at $1.5\,\mu m$. Microstructures ($50\times 50\,\mu m^2$) of 1D double period metallic gratings are obtained by a lift-off process~\cite{supp}: gratings are first defined by electronic beam lithography followed by a $50\,nm$ gold deposition and development to etch away the resist or lithography. Scanning electronic microscopy (SEM) images of a typical structure is shown in Fig.~2(a), highlighting the unit cell of the double period design [inset of Fig.~2(a)]. 
\begin{figure}[htb]
\begin{center}
\includegraphics[width=8.5cm]{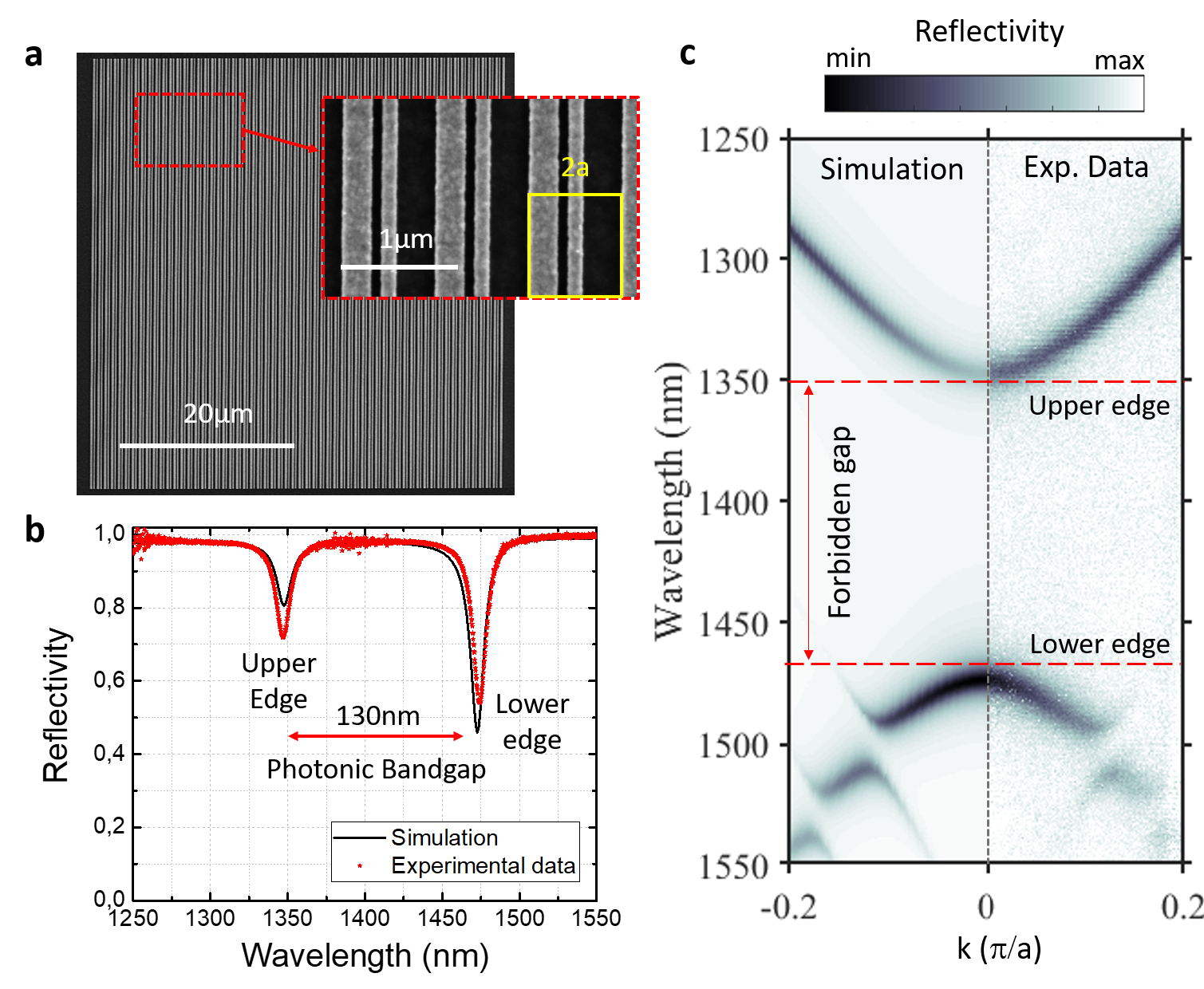}
\end{center}
\caption{\label{fig2}{(a) SEM image of a double period structure. Inset is a zoom to highlight the structure details on a single unit cell. (b) Reflectivity spectrum at normal incidence of structure with $a=400\,nm$, $w=260\,nm$, $f=0.55$ and $s=70\,nm$. The red dots are experimental data while the black line is the RCWA calculation. (c) Dispersion diagram obtained by angle-resolved reflectivity spectra. The right (left) panel corresponds to experimental data (RCWA calculation).}}
\end{figure}
\begin{figure}[htb]
\begin{center}
\includegraphics[width=8.5cm]{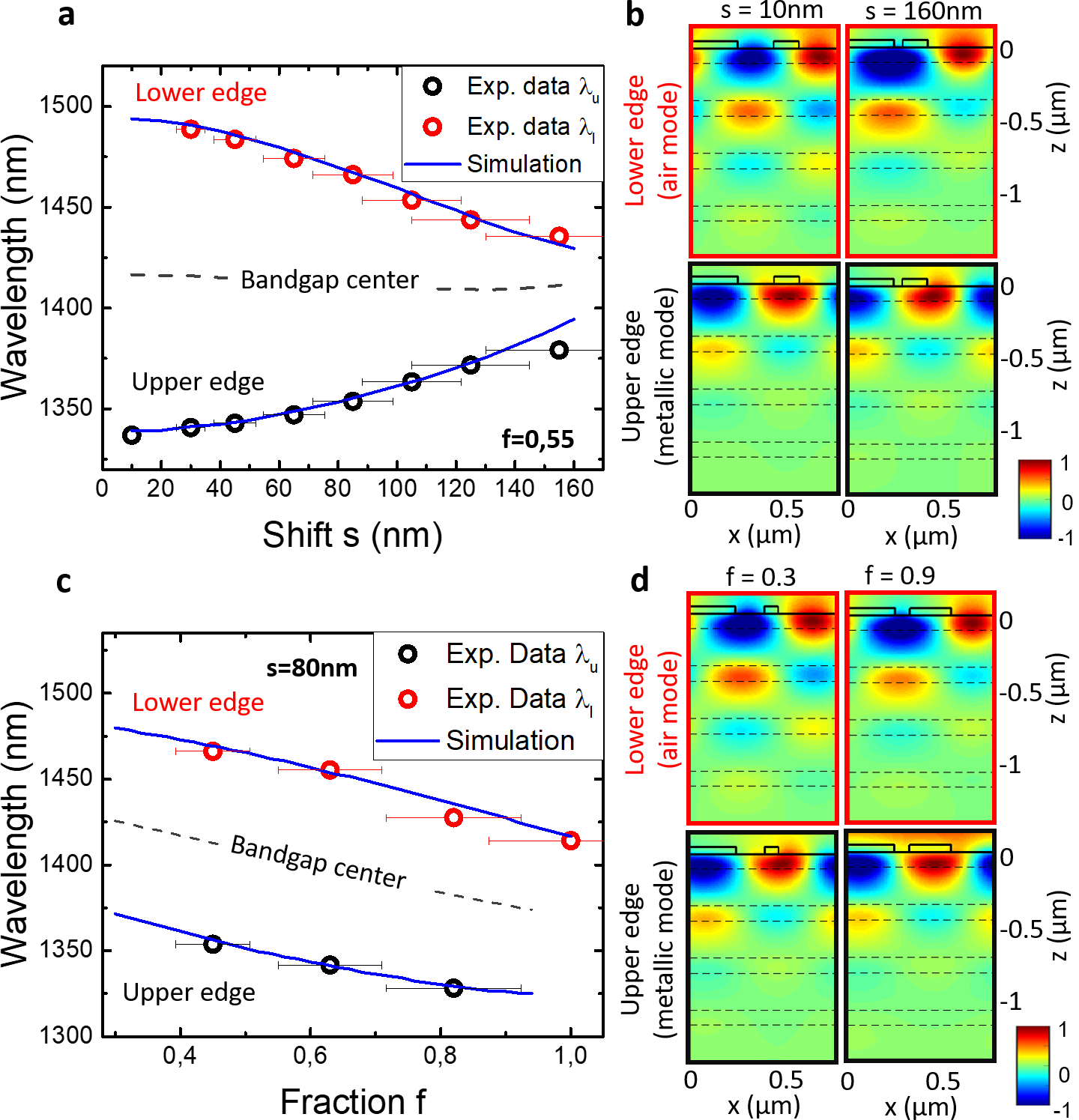}
\end{center}
\caption{\label{fig3}{Bandgap tailoring by playing with the double period design parameters s (shift) and fraction f
(a,c) Band-edges position while varying (a) the shift parameter while fixing the shrinking parameter at $f=0.55$, (c) the shrinking parameter while fixing the shift parameter at $s=80\,nm$. The blue line is the results of RCWA calculation. (b,d) the field distribution corresponding to extreme cases of (a) and (c).}}
\end{figure}

Reflectivity spectrum at normal incidence of structure with $a=400\,nm$, $w=260\,nm$ and $f=0.55$ is presented in Fig.~2(b). Two dips corresponding to the two band-edges, separated by a bandgap of $130\,nm$, are experimentally observed, and perfectly reproduced by numerical simulation with Rigorous coupled-wave analysis (RCWA) method. To demonstrate that these dips correspond to the band-edges of two Tamm modes with opposite curvature, we visualize the band diagram in the momentum space by performing angle-resolved micro-reflectivity measurements~\cite{supp}. The experimental band diagram, shown in the right panel of Fig.~2(c), evidences clearly the two band-edges at $k_x=0$ and a complete photonic bandgap of $130\,nm$, centred at $1412\,nm$. These data are amazingly reproduced by RCWA calculations [left panel of Fig.~2(c)]. In the following, we will show how the bandgap demonstrated previously can be tailored on-demand by playing with the shift parameter $s$ and the shrinking parameter $f$: 

\underline{\textit{Bangap-size tuning:}} Figure~3(a) reports experimental measurements and RCWA calculations of the band-edges positions when varying the shift parameter $s$, but keeping fix the shrinking parameter $f=0.55$. These results show clearly that just simply by changing the shift parameter, the bandgap-size can be tuned from $30\,nm$ to $150\,nm$ (i.e. $500\%$ of variation) without any moving of its central wavelength. This effect is explained by a confinement reduction of the upper edge mode, together with a confinement enhancement of the lower edge state when increasing $s$. Indeed, the electric field distribution of each each band-edge mode is presented in Fig.~3(b) for two extreme values of the shift parameter ($s=10\,nm$ and $s=160\,nm$). It shows that for $s=10\,nm$, the field distribution of each mode is almost the same as the case of simple period design [Fig.~1(c)]: the upper edge is a ``metal'' mode with field confined beneath the metal stripes; and the lower edge is an ``air'' mode with field confined beneath the air grooves. However, for $s=160\,nm$,  a big part of the field of the upper edge is turned to be located beneath the air grooves; and a big part of the field of the lower edge is turned to be located beneath the metal stripes. As consequence, the confinement of the upper edge mode is reduced while the one of the lower edge mode is enhanced. 
\begin{figure*}[htb]
\begin{center}
\includegraphics[width=15cm]{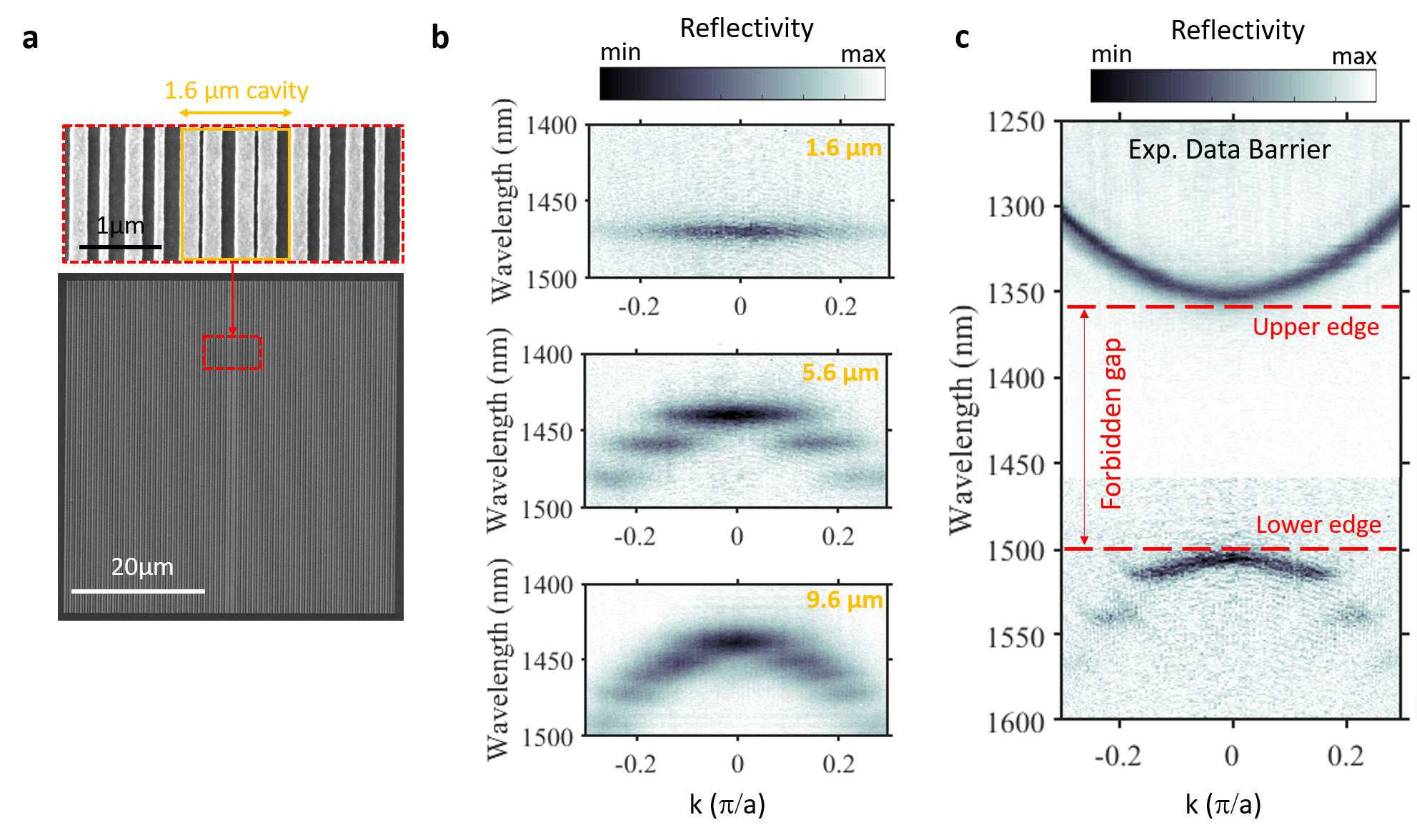}
\end{center}
\caption{\label{fig4}{(a) SEM images of the 2 unit cells cavity. (b,c) Experimental measurements of the angle-resolved micro-reflectivity experiment performing on  (b) three cavities of size of 2 unit cells ($1.6\,mum$), 7 unit cells ($5.6\,mum$) and 12 unit cells ($9.6\,mum$), (c) the barrier outside of the cavities.}}
\end{figure*}

\underline{\textit{Bandgap-center tuning:}} Figure~3(c) reports experimental measurements and RCWA calculations of the band-edges positions when varying the shrinking parameter $f$, but keeping fix the shrinking parameter $s=80\,nm$. Contrary to the previous observation,  these results show that the bandgap-size stays roughly at the same value (i.e.$\sim\,120\,nm$) but its central wavelength can be freely moved within $50\,nm$ (i.e. $42\%$ of the bandgap), just simply by changing the shrinking parameter. This effect is explained by the a confinement enhancement of both band-edge modes when decreasing the shrinking parameter $f$. Indeed, the electric field distribution of each each band-edge mode is presented in Fig.~3(d) for two extreme values of the shrinking parameter ($f=0.3$ and $f=0.9\,nm$). On one hand, for both case, the upper-edge is always a ``metal'' mode and the lower-edge is always an ``air'' mode. This probably explains why the bandgap-size stays at the same values for different values of $f$ . On the other hand, when the second gold stripe is shrunk, there is more air space for both modes to leak, leading to a confinement reduction for both of them. This explains the redshift of the bandgap-center when decreasing $f$ (i.e. increasing the shrinking).

Finally, we demonstrate the ultimate bandgap engineering: implementation of photonic heterostructures to make photonic crystal defect cavities~\cite{Song2005,Istrate2006,Ferrier2008}. These heterostructures are constructed by introducing a local geometrical modification (i.e. defect) on a photonic crystal so that photons  with a specific wavelength are allowed in the defect but fall into the bandgap of the outside structure (i.e. barrier)~\cite{Istrate2006}. This leads to a confinement of photons within the defect, namely cavity confinement. In our heterostructure design, we consider a photonic barrier of parameters $a=400\,nm$, $w=250\,nm$, $f=0.5\,nm$ and $s=60\,nm$. The defect cavity has exactly the same parameters as the barrier, except the shrinking parameter $f=1$. According to results shown in Fig.~3(c), the bandgap-size of the barrier and the defect are the same, but the bandgap-center of the defect is redshifted with respect to the one of the barrier. This results in a confinement of the lower edge mode within the bandgap of the barrier structure. To evidence this confinement, three cavities of different sizes have been fabricated. The corresponding cavity sizes are 2 unit cells ($1.6\mu m$), 7 unit cells ($5.6\mu m$) and 12 unit cells ($9.6\mu m$). Figure~4(a) presents SEM images of the  2 unit cells cavity, illustrating the fabrication quality of our samples. Angle-resolved micro-reflectivity experiment is performed on each cavity and also on the barrier. The dispersion diagram obtained from the experimental measurements are reported in Fig.~4(b) and (c). These results show clearly, for all cavities, the apparition of discrete Tamm modes inside the barrier bandgap, with mode spacing increases when reducing the cavity size. Interestingly, unlike the case of confinement induced by metalic patch~\cite{Gazzano2011,Symonds2013} where the initial Tamm mode has positive effective mass, the Tamm mode that is confined within our cavities is the lower edge mode of negative effective mass. That explains why the fundamental mode of our cavities is the one at highest energy. We highlight that in the case of the smallest cavity (i.e. 2 unit cells size), the confinement is strong enough to push away high-order modes out side of the barrier bandgap, and only the fundamental mode is observed. 

In conclusion, we demonstrate for the first time the band diagram engineering of optical Tamm modes by periodically patterning the metal layer in the photonic crystal regime ($a/\lambda\ll1$). By adopting a double period design, we evidenced experimentally a complete photonic bandgap  up to $150\,nm$ in the telecom wavelength range. Moreover, our design makes it possible to tailor on-demand, and independently, the band-gap size and its spectral position. Experimental results highlight a continous tuning of the bandgap size from $30\,nm$ to $150\,nm$, and of the bandgap center within $50\,nm$. All of our experimental data are perfectly reproduced by numerical calculations. Moreover, by introducing a defect cavity within our Tamm plasmon photonic crystal, an ultimate cavity of $1.6\mu m$ supporting a single highly confined Tamm mode is experimentally demonstrated. We highlight that all 1D photonic crystal concepts developed in this work can be naturally applied for the case of 2D photonic crystal, with the possibility of even more degrees of freedom for bandgap engineering. Our results suggests the possibility to engineer novel band structures such as Dirac cones~\cite{Huang2011,Li2015c,Zhen2015,Nguyen2018} and flat bands~\cite{Vicencio2015,Mukherjee2015,Nguyen2018} with surface modes in hybrid metal/dielectric structures. This would pave the way towards contemporary topics of optics research such as topological photonics~\cite{Lu2014}, metamaterials~\cite{Shalaev2007,Huang2011,Segal2015,Li2015c}, coupling with 2D materials~\cite{Xia2014,Noori2016}, non-hermitian physics and parity-time symmetry ~\cite{Zhen2015,Feng2017,El-Ganainy2018}.

The authors would like to thank the staff from the NanoLyon Technical Platform for helping and supporting in all nanofabrication processes. This work is partly supported by the French National Research Agency (ANR) under the projects NEHMESIS.

\bibliography{RefTAMM}

\end{document}